\documentclass[%
 reprint,
superscriptaddress,
 amsmath,
 amssymb,
 aps,
]{revtex4-2}

\usepackage{graphicx}
\usepackage{dcolumn}
\usepackage{bm,amsmath}
\usepackage{hyperref}

\hypersetup{
    colorlinks,
    citecolor=blue,    filecolor=blue,
    linkcolor=blue,    urlcolor=blue
}

\DeclareMathOperator\erf{erf}

\begin{document}

\title{Advanced Shaping of Quasi-Bessel Beams for High-Intensity Applications}

\author{J\'er\^ome Touguet}
\affiliation{Laboratoire d’Optique Appliqu\'ee, ENSTA, CNRS, Ecole polytechnique, Institut Polytechnique de Paris, 828 Bd des Mar\'echaux, 91762 Palaiseau, France}
\author{Igor Andriyash}
\affiliation{Laboratoire d’Optique Appliqu\'ee, ENSTA, CNRS, Ecole polytechnique, Institut Polytechnique de Paris, 828 Bd des Mar\'echaux, 91762 Palaiseau, France}
\author{Ronan Lahaye}
\affiliation{Laboratoire d’Optique Appliqu\'ee, ENSTA, CNRS, Ecole polytechnique, Institut Polytechnique de Paris, 828 Bd des Mar\'echaux, 91762 Palaiseau, France}
\author{Guillaume Chapelant}
\affiliation{Laboratoire d’Optique Appliqu\'ee, ENSTA, CNRS, Ecole polytechnique, Institut Polytechnique de Paris, 828 Bd des Mar\'echaux, 91762 Palaiseau, France}
\affiliation{THALES AVS - MIS, 2 Rue Marcel Dassault, V´elizy-Villacoublay, 78140, France}
\author{Julien Gautier}
\affiliation{Laboratoire d’Optique Appliqu\'ee, ENSTA, CNRS, Ecole polytechnique, Institut Polytechnique de Paris, 828 Bd des Mar\'echaux, 91762 Palaiseau, France}
\author{Lucas Rovige}
\affiliation{Laboratoire d’Optique Appliqu\'ee, ENSTA, CNRS, Ecole polytechnique, Institut Polytechnique de Paris, 828 Bd des Mar\'echaux, 91762 Palaiseau, France}
\author{C\'edric Thaury}
 \email{cedric.thaury@ensta.fr}
\affiliation{Laboratoire d’Optique Appliqu\'ee, ENSTA, CNRS, Ecole polytechnique, Institut Polytechnique de Paris, 828 Bd des Mar\'echaux, 91762 Palaiseau, France}

\begin{abstract}
Quasi-Bessel beams produced by axiparabolas are increasingly used in high-intensity laser applications, yet their longitudinal profiles exhibit unwanted oscillations that limit their effectiveness. Here we identify the physical origin of these distortions and develop a general strategy to control the on-axis intensity of extended focal lines. By combining analytical insight with numerical and experimental validation, we show how both smooth and sharply structured longitudinal profiles can be reliably produced. This establishes a robust framework for tailoring quasi-Bessel beams in regimes relevant to laser–plasma acceleration, advanced photon sources, and other high-field applications.
\end{abstract}

\maketitle
\section{Introduction }
Non-diffractive light beams have attracted significant attention for their ability to maintain a narrow intensity profile over long propagation distances, making them valuable for a wide range of applications in optics and photonics. Among the various types of beams exhibiting this behavior, Bessel beams are likely the most extensively studied. They are characterized by a concentric ring pattern described by a Bessel function of the first kind. In theory, an ideal Bessel beam could remain perfectly focused over an infinite distance, but such a beam would require infinite energy to generate. In practice, various experimental techniques allow the generation of finite-energy approximations that retain their shape over a limited propagation range, making them effectively quasi-non-diffractive ~\cite{PhysRevLett.58.1499, 2005ConPh..46...15M}.

Several recent studies have shown that quasi-Bessel beams are highly effective at very high laser intensities. For example, they can be used to generate plasma waveguides~\cite{PhysRevLett.71.2409,Oubrerie2022} or to enhance the energy of electron beams produced by laser-plasma acceleration by suppressing the dephasing between the driver and the accelerated beam~\cite{clement, PhysRevLett.124.134802,Malka_Nature_comms}. Most of these applications benefit from, or even require, optical components capable of withstanding high laser intensities while preserving the ultra-short duration of the laser pulse. Consequently, the techniques based on reflective optics are preferred for generating Bessel beams in such applications. To address this need, the axiparabola~\cite{Smartsev:19,FAN2019124342} was introduced as a reflective analogue of the axilens~\cite{1991OptL...16..523D}. Unlike axilenses, axiparabolas are achromatic and feature a high damage threshold, making them particularly well suited for ultra-intense, broadband laser pulses. Both axiparabolas and axilenses optimize laser energy usage by allowing the position and length of the non-diffractive focal line to be precisely matched to the interaction region.

The utility of these optics extends beyond controlling the position or length of the non-diffractive zone; their design also enables phase optimization to tailor the longitudinal intensity profile. For instance, it is possible to achieve a uniform intensity along the focal line or a progressively increasing intensity, which can be used to maintain constant enclosed energy in the first Bessel ring~\cite{2022JOpt...24d5503O}.
However, regardless of the peak intensity, longitudinal profile, or generation technique, the quasi-Bessel beam’s finite radial extent inevitably restricts the length of its non-diffractive region.
This mathematically corresponds to multiplying the transverse laser distribution by an amplitude gate applied far from the focus. Depending on the shape of this gate function, the truncation can introduce modulations along the non-diffractive focal line~\cite{2005ConPh..46...15M}. 
While controlled modulations can be beneficial for certain applications, such as enhancing X-ray radiation in laser-plasma accelerators~\cite{PhysRevLett.112.134803,Tomkus2020}, unwanted modulations severely disrupt most applications. In particular, in laser-plasma accelerators, these modulations induce fluctuations in the length of the accelerating cavity, causing unwanted multiple injections during cavity expansion and charge losses during contraction. This completely undermines  precise control of the injected charge profile, thus preventing the production of high-quality electron beams.

In this study, we first analytically identify the origin of unwanted modulations in the focal line. We then demonstrate, both numerically and experimentally, that these oscillations can be suppressed using either an amplitude mask or an appropriate phase profile. Finally, we show how controlled modulations can be intentionally reintroduced into the intensity profile to meet the requirements of specific applications.

\section{Analytical description }
\label{sec:analytical}
To illustrate the origin of the oscillations along the focal line, we consider, for simplicity, a radially symmetric and monochromatic top-hat beam with intensity  $I_0$ and an angular frequency $\omega$, incident on an axiparabola mirror. We also allow for the possibility that the mirror includes a central hole of radius $r_h$, as required in certain applications~\cite{PhysRevLett.71.2409,Oubrerie2022}. The mirror imprints a radial phase $\phi(r)$ on the beam, thereby focusing rays originating from different radial positions $r$ to distinct axial positions $z(r)$, where $z=0$ corresponds to the mirror surface.  In the Fresnel diffraction regime, the electric field $E$ along the focal line is given by
\begin{align}
    \label{eq:fresnel}
    E(r_{\zeta},z)&=-i\frac{k E_0}{z}e^{ik\left(z+\frac{r_{\zeta}^2}{2z}\right)}\int_{r_h}^Rrdre^{i\Psi(r)}J_0\left(k\frac{r_{\zeta}r}{z}\right)\text{,}
\end{align}
where  $r_{\zeta}$  is the radial position, $k=\omega/c$, $\Psi(r) = kr^2/2z-\phi(r)$ and $R$ is the beam radius on the mirror surface~\cite{2022JOpt...24d5503O}.

Since $e^{i\Psi(r)}$  is a rapidly oscillating function, the integral can be approximated using the stationary phase method. This method assumes that contributions to the integral cancel out due to destructive interference, except near stationary phase points, where $\Psi'(r)=0$, allowing the integral to be approximated by its local behavior around these points. To this end, we perform a Taylor expansion of the phase around the stationary point $r_s(z)$, where $\Psi'(r_s)=0$. The on-axis field can then be expressed as
\begin{align}
    E(0,z)&=-i\frac{k r_s}{z}E_0 e^{ikz +i\Psi(r_s)}\int_{r_h}^Rdre^{i\Psi''(r_s)(r-r_s)^2/2}\text{.}
    \label{eq:stat_approx}
\end{align}
 Note that, in an optical geometry approach, $r_s(z)$ would represent the radius on the axiparabola mirror of rays that intersect the optical axis at $z$. 
 
 The integration of Eq.~(\ref{eq:stat_approx}) reduces to evaluating an integral of the form $\int_a^b e^{i\alpha r^2} dr\text{,}$ with $a  <0 <b$  which can be approximated as
\begin{align}\int_a^b e^{i\alpha r^2} dr = \sqrt{\frac{\pi}{\alpha}}e^{i\pi/4} + i\left(\frac{e^{i\alpha a^2}}{2\alpha a} -\frac{e^{i\alpha b^2}}{2\alpha b}\right)+\mathcal{O}\left(\frac{1}{\alpha^2}\right)\text{.}
\label{eq:int_expr2}
\end{align}
In general, in the stationary phase approach, only the first term is retained, so the on-axis field simplifies to 
\begin{align}
\label{eq:field_stat}
    E(z)=-i E_0\frac{k r_s}{z}\sqrt{\frac{2\pi}{\Psi''(r_s)}}e^{i\left( k z +\Psi(r_s)+\pi/4\right)}\text{.}
\end{align}

Equation~(\ref{eq:field_stat}) provides a good estimate along most of the focal line, but near the edges ($r_s \rightarrow r_h$ or $r_s \rightarrow R$), the first term alone is insufficient, and the next term has to be included. Specifically, near the lower bound ($r_s \rightarrow r_h$) the term $\propto 1/a$ must be retained, while near the upper bound ($r_s \rightarrow R$) the term $\propto 1/b$ must be retained. This yields
\begin{align}
    E(z)&\underset{r_s\rightarrow r_h}{\approx} A \left(ir_s e^{i\pi/4}\sqrt{\frac{2\pi}{\Psi''(r_s)}}+  \frac{e^{i (r_h-r_s)^2\Psi''(r_s)/2)}}{\Psi''(r_s)(1-r_h/r_s)}\right)\nonumber\\
     E(z)&\underset{r_s\rightarrow R}{\approx} A \left(ir_s e^{i\pi/4}\sqrt{\frac{2\pi}{\Psi''(r_s)}}+  \frac{e^{i (R-r_s)^2\Psi''(r_s)/2)}}{\Psi''(r_s)(R/r_s-1)}\right)\text{,}
     \label{eq:field_model}
\end{align}
where $A=-(E_0 k /z)\exp{[i k z +i\Psi(r_s)]}$. Compared to Eq.~(\ref{eq:field_stat}), an additional complex exponential term appears, causing intensity oscillations at the beginning and end of the focal line. The contrast of these oscillations tends to zero when the first term is much larger than the second one. For instance, when $r_h = 0$, the profile is expected to exhibit significant oscillations unless 
\begin{align}2\pi |\Psi''| \gg 1/r_s^{2}\mathrm{,}
\label{eq:cond}
\end{align}
a condition that is generally not satisfied, since the phase typically varies slowly in the vicinity of $r=0$.
\begin{figure}
    \centering
    \includegraphics[width=\linewidth]{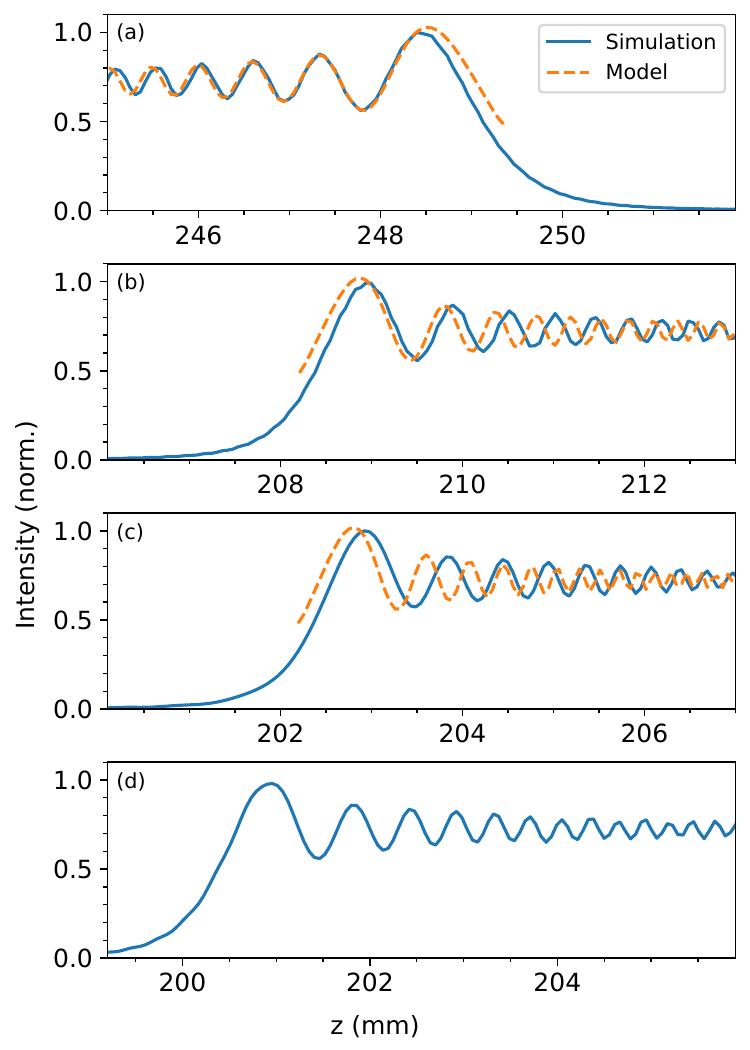}
    \caption{Intensity along the optical axis at the end (a) and at the beginning (b, c, d) of the focal line for a hole radius of $r_h = 20$~mm (b), $r_h = 10$~mm (c), and $r_h = 0$~mm (d). The blue curves were computed with \texttt{Axiprop}, while the orange curves were obtained using Eq.~(\ref{eq:field_model}).}
    \label{fig:oscil_theo}
\end{figure}

To validate these theoretical predictions, we used the \texttt{Axiprop} optical propagation Python library~\cite{axiprop}, which relies on discrete Hankel and Fourier transforms. The simulations consider a beam of radius $R = 50$~mm and an axiparabola mirror designed to produce a focal line of constant intensity and length $\delta = 50$~mm, starting at $z = f_0 = 200$~mm (yielding, to first order, $r_s^2 / R^2 = (z - f_0) / \delta$).

Figure~\ref{fig:oscil_theo} shows the intensity along the optical axis at the end (a) and at the beginning (b) of the focal line for a central hole radius $r_h = 20$~mm. The agreement between the intensity predicted by Eq.~(\ref{eq:field_model}) (orange curves) and the exact numerical result (blue curves) is excellent at the end of the focal line and fair at the beginning. This agreement degrades as the hole radius $r_h$ decreases, as illustrated in Fig.~\ref{fig:oscil_theo}(c). This behavior occurs because the value of $\Psi''$ becomes smaller at low $r$, reducing the accuracy of the Taylor expansion in Eq.~(\ref{eq:stat_approx}). Without a central hole, $\Psi$ varies too slowly for the stationary-phase approximation to hold, and the model breaks down. Nevertheless, the intensity profile still exhibits similar oscillations, as shown in Fig.~\ref{fig:oscil_theo}(d), indicating that they do not originate from the hole but from edge effects associated with the sharp onset and termination of the focal line.

Despite the model’s limitations for small $r_h$, the strong agreement obtained for sufficiently large $\Psi''$ provides clear evidence that the truncation of the focal line is the origin of these oscillations. A direct implication of this analysis is that oscillations can be effectively suppressed by avoiding abrupt intensity transitions, which can be achieved by optimizing either the amplitude or the phase of the incident beam.

\section{Suppression of the oscillations with amplitude modulation}
\label{sec:ampl}
\subsection{Model}
\label{sec:ampl_theo}
\begin{figure}
    \centering
    \includegraphics[width=\linewidth]{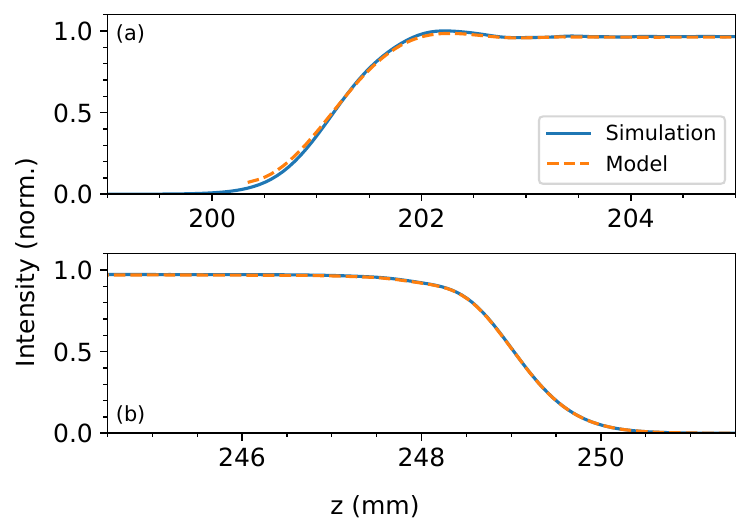}
    \caption{Intensity along the optical axis at the beginning (a) and at the end (b) of the focal line for a super-Gaussian laser pulse with a Gaussian central hole, as defined by Eq.~(\ref{eq:def_las}).}
    \label{fig:supr_ampl_theo}
\end{figure}

We first consider the use of an amplitude mask. As demonstrated in Section~\ref{sec:analytical}, the oscillations originate from the sharp cutoff of the Fresnel integral. Consequently, we expect to suppress these oscillations by illuminating the axiparabola with a smoother intensity profile~\cite{FAN2019124342}.
To validate this approach, we now replace the top-hat profile of the incident laser by a smoother super-Gaussian of the order $n$, and substitute the sharp central hole with a Gaussian profile:
\begin{align}
E_i(r)= E_0 e^{-(r/R)^{2n}}\begin{cases}
           e^{-{(r-r_{h})^2}/{L_{h}^2}} \quad &\text{if } \, r\leq r_{h} \\
           1\quad &\text{if } \, r>r_{h}\\
     \end{cases} 
     \label{eq:def_las}
\end{align}
where $L_h$ is the scale length of the Gaussian hole. The electric field along the focal line is then simply obtained by introducing $E_i(r)$ into Eq.~(\ref{eq:stat_approx}),as 
\begin{align}
    E(z)&=-i\frac{k r_s}{z} e^{ikz +i\Psi(r_s)}\int_{0}^RdrE_i(r)e^{i\Psi''(r_s)(r-r_s)^2/2}\text{.}
    \label{eq:stat_approx_prof}
\end{align}
The numerical integration of Eq.~(\ref{eq:stat_approx_prof}) is plotted in Fig.~\ref{fig:supr_ampl_theo}, where it is compared to the results of an \texttt{Axiprop} simulation. The parameters used are identical to those in Sec.~\ref{sec:analytical}, except for $r_h=10$~mm, $L_h=4$~mm, and $n=100$. We observe an excellent agreement between the model and the simulations, as well as efficient suppression of the intensity of oscillations.

\subsection{Experiment}
\label{sec:ampl_exp}
To verify the theoretical predictions, we performed an experiment using a He–Ne laser with a central wavelength of $\lambda = 633$~nm. The top-hat beam was directed onto an axiparabola designed to produce 
a constant-intensity line, characterized by $f_0 = 200$~mm and $\delta = 30$~mm, for an input beam diameter of $R_m = 50.8$~mm.
Amplitude modulation was applied using a Holoeye LC2012 spatial light modulator (SLM) with a resolution of 1024 × 768 pixels and a pixel pitch of 36~$\mu$m. The Gaussian beam delivered by the He–Ne laser was expanded before the SLM to achieve a beam waist of 40~mm, and then apertured to a diameter of 27.6~mm to match the active area of the modulator. After the SLM, the beam was expanded by a factor of 2.5, resulting in a truncated Gaussian beam with a waist of approximately 100~mm and an aperture of 69~mm on the axiparabola surface.
The focal region was imaged onto a CMOS camera through a microscope objective. The imaging system (microscope and camera) was mounted on a motorized translation stage, allowing the measurement of the axial evolution of the on-axis intensity distribution.

\begin{figure}
    \centering
    \includegraphics[width=\linewidth]{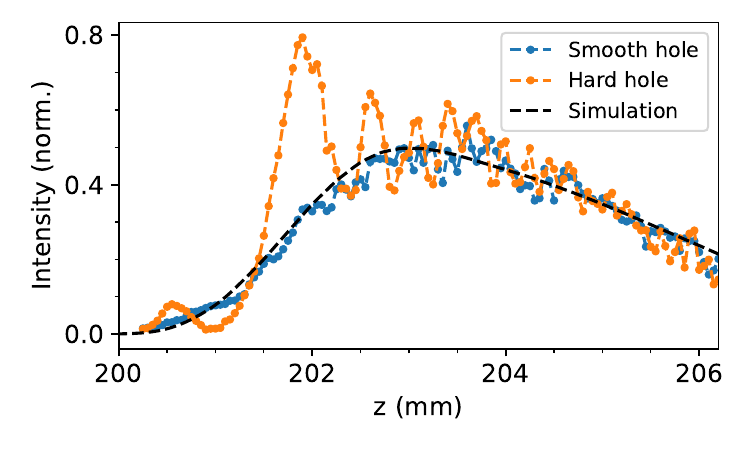}
\caption{Experimental on-axis intensity profiles obtained with a sharp hole (orange) and a Gaussian hole (blue) with $r_h = 11.6$~mm and $r_h = 16.75$~mm respectively. Theoretical prediction from simulation is shown as dashed lines.}    \label{fig:exp_ampl}
\end{figure}

Figure~\ref{fig:exp_ampl} presents the recorded on-axis intensity profiles for both a sharp and a smooth hole, as defined by Eq.(\ref{eq:def_las}), for $r_h = 11.6$~mm and $r_h = 16.75$~mm, respectively, with $L_h = 5.15$ mm. Each data point corresponds to the average of 10 images acquired with an exposure time of 200~ms. Similar to Figs.~\ref{fig:oscil_theo} and~\ref{fig:supr_ampl_theo}, pronounced oscillations are observed when using a sharp hole, while they are well suppressed with the Gaussian hole. A noticeable difference is observed in the intensity profile: instead of reaching the expected plateau after the hole, the intensity decreases. This behavior can be attributed to a trefoil aberration generated by the axiparabola, which distorts the focal intensity distribution. Introducing a trefoil aberration ($\Phi_{\text{trefoil}}(r,\theta)=2\pi\,a_3\,\cos\left(3\theta\left(\dfrac{r}{R}\right)^3\right)$ in Eq.~(\ref{eq:fresnel}) modifies the amplitude distribution by the factor
\begin{align}
    J_0\left(2\pi a_{3}\left(\frac{z-f_0}{\delta}\right)^{3/2}\right),
    \label{eq:trefoil}
    \end{align}
where \(a_3\) is the trefoil coefficient expressed in units of the wave number. 
The simulated curve shown in Fig.~\ref{fig:exp_ampl} implements the amplitude modulation described by Eq.~\ref{eq:trefoil}, resulting in good agreement with the experimental data, with a measured trefoil coefficient of \(a_3 = 2.2\).Importantly, this reduction in intensity could be mitigated by using deformable mirrors to correct for the aberration.

\section{Suppression of the oscillations with phase only optics}
\subsection{Model}
We showed in Sec.~\ref{sec:ampl} that oscillations can be efficiently suppressed by applying amplitude modulation on the incoming beam. Although oscillations originating from the outer edge of the beam are often naturally suppressed, since real beams are naturally super-Gaussian, the suppression of oscillations originating from the central part of the beam is more challenging. Indeed, producing a smooth intensity dip at the center of the incoming beam would require absorbing a significant fraction of the incident energy, which could damage the optical components used for that purpose. Suppressing these oscillations by acting only on the phase of the beam, through the use of a focusing optic, a spatial light modulator, or a phase mask would therefore be highly desirable. In this section, we investigate the conditions under which such phase-based suppression can be achieved.

Different strategies can be adopted to avoid oscillations at the beginning of the focal line. Since these oscillations originate from abrupt variations in the desired intensity profile, one approach is to define a longitudinal intensity distribution that smoothly rises from low values before transitioning into a constant plateau, similar to what can be obtained through amplitude shaping, and then compute the phase that generates this profile. An alternative approach would consist in using an iterative algorithm to determine the optimal phase under given constraints~\cite{Cizmar:09}.

Here, we adopt the simplest approach and start directly from a given phase profile, which we modify to suppress the oscillations. As discussed in Sec.~\ref{sec:analytical}, oscillations arise when the second derivative of the radial beam phase before focus is too small to blur interferences between different regions of the beam. Consequently, oscillations can be mitigated by artificially increasing $\Phi''(r)$ for any $r < r_0$, where $r_0$ is an arbitrary radius. For example, one can simply replace the beam phase for $r < r_0$ by
\begin{equation}
\Phi(r) = \Phi(r_0) + \Phi'(r_0)\,(r - r_0) + \frac{1}{2} \Phi''(r_0)\,(r - r_0)^2,
\label{eq:phase_mod}
\end{equation}
so that $\Phi''(r)$ remains constant for $r < r_0$.

\begin{figure}
    \centering
    \includegraphics[width=\linewidth]{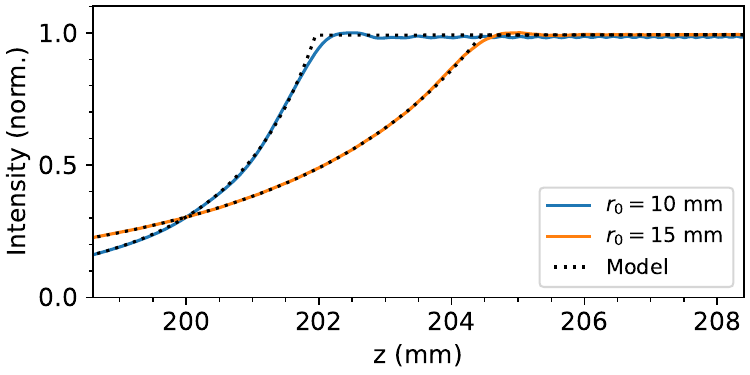}
    \caption{On-axis intensity profiles obtained from \texttt{Axiprop} simulations using the modified phase approach. Profiles are shown for $r_0 = 10$~mm (blue line) and $r_0 = 15$~mm (orange  line). the dotted lines correspond to the profiles obtained from Eq.~\ref{eq:field_stat}.}
    \label{fig:supr_phase_theo}
\end{figure}

This approach was validated through two \texttt{Axiprop} simulations performed for $r_0 = 10$~mm and $r_0 = 15$~mm, using the same axiparabola parameters as in Secs.~\ref{sec:analytical} and~\ref{sec:ampl_theo}. As shown in Fig~\ref{fig:supr_phase_theo}, oscillations are strongly damped for $r_0 = 10$~mm, similar to 
amplitude shaping, and almost completely suppressed for $r_0 = 15$~mm. 
Since Eq.~(\ref{eq:cond}) is satisfied, the on-axis intensity  can be computed directly from Eq.~(\ref{eq:field_stat}). The resulting profiles, plotted in Fig.~\ref{fig:supr_phase_theo}, show excellent agreement with the numerical simulations. Furthermore, the intensity evolution within the rising ramp can be obtained analytically by deriving $r_s(z)$ for the phase defined in Eq.~(\ref{eq:phase_mod}), leading to $$I(z) \propto \frac{z}{\left(k_0 - z \Phi''(r_0) \right)^3}.$$ This expression shows that the intensity profile exhibits a long leading edge, vanishing only as $z$ approaches zero.
In practice, the shape and extent of this ramp could be readily adjusted to meet the requirements of a specific experiment by employing a more elaborate phase profile.

\subsection{Experiment}
As in Sec.~\ref{sec:ampl_exp} for the amplitude-based approach, we now turn to an experimental investigation to validate the predictions of the phase-based method. The experimental configuration is identical to the previous setup, except for one important modification: because the phase delay achievable with a single SLM is limited, we employed two SLMs in combination. In addition, we used a laser diode with a central wavelength of $\lambda = 405$~nm.\begin{figure}
    \centering
    \includegraphics[width=\linewidth]{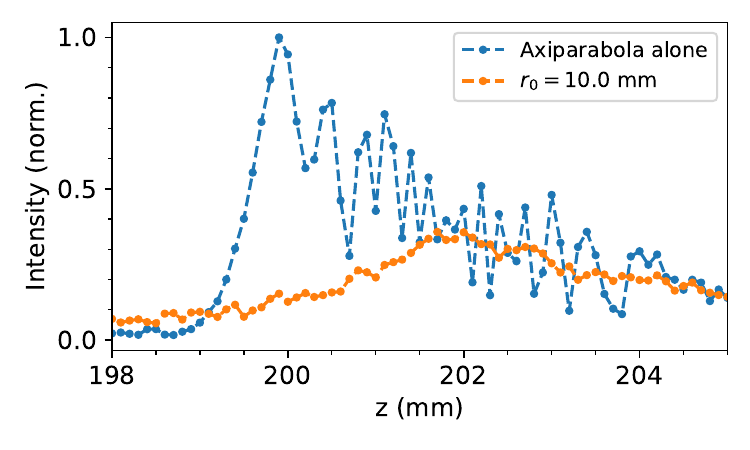}
    \caption{Experimental on-axis intensity profiles obtained without modulation (blue) and a phase modulation (orange) with $r_h = 10$~mm.}
    \label{fig:phase_modulation}
\end{figure}

Figure~\ref{fig:phase_modulation} shows the recorded intensity profiles for the unmodulated beam and for the beam shaped with a phase modulation of $r_0 = 10$~mm. Each data point corresponds to the average of 20 images acquired with an exposure time of 100~ms. We observe that the oscillations are strongly damped, with a long rising ramp, in good agreement with the simulations.

A notable difference, however, is the decrease in intensity along the line, whereas a nearly constant intensity was expected. 
This discrepancy likely results from the use of two SLMs, which introduces noticeable aberrations, further aggravated by one device’s partial pre-existing deterioration.

\section{Complex intensity profiles }
\subsection{Resolution limit}
\label{sec:lmt_res}
\begin{figure}
    \centering
    \includegraphics[width=\linewidth]{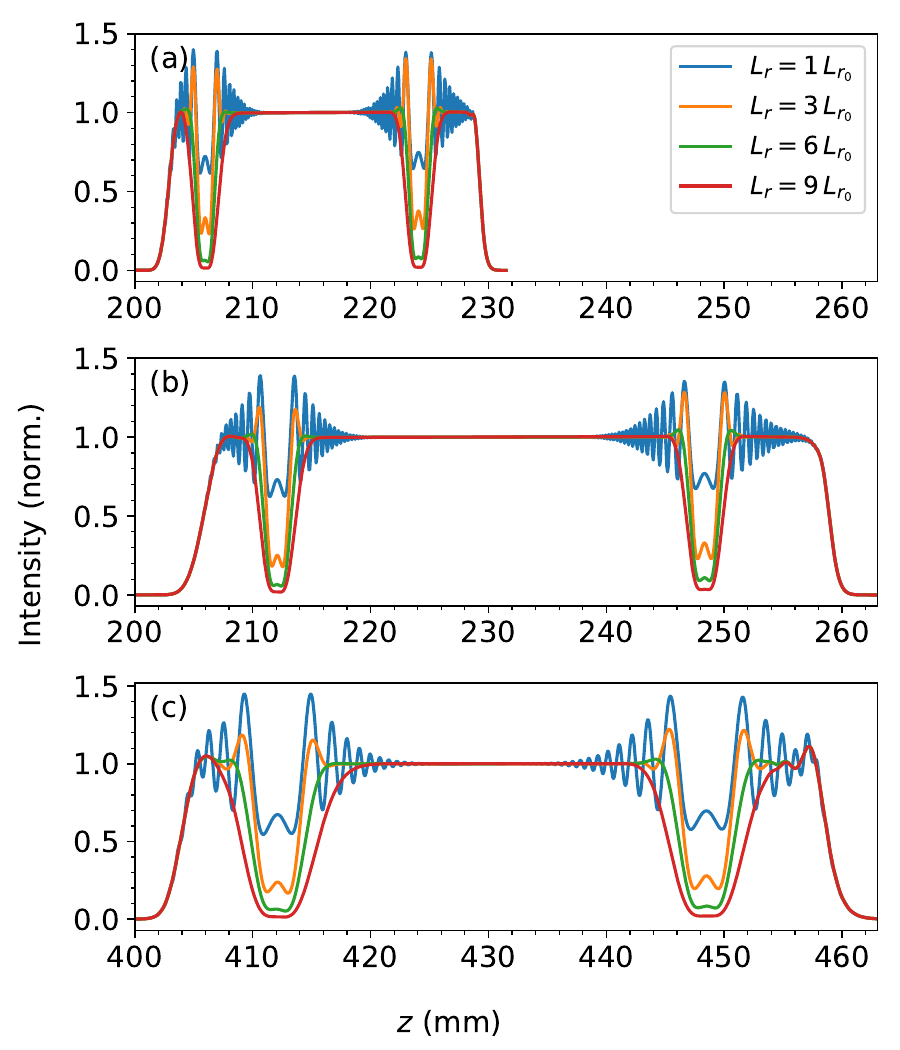}
    \caption{On-axis intensity profiles obtained from simulations for different axiparabola parameters. (a) $f_0 = 200$ mm and $\delta = 30$ mm. (b) $f_0 = 200$ mm and $\delta = 60$ mm. (c) $f_0 = 400$ mm and $\delta = 60$ mm. The curves correspond to different values of $L_r \in \{1,3,6,9\}L_{r_0}$}
    \label{fig:res}
\end{figure}

In the previous sections, we demonstrated the production of extended focal lines free of unwanted modulations. However, it may sometimes be desirable to produce controlled modulations. From Sec.~\ref{sec:analytical}, we know that sharp transitions lead to unwanted oscillations; therefore, we cannot generate structures of arbitrarily small size. To estimate the achievable resolution, we can use Eq.~(\ref{eq:int_expr2}) with $b=\infty$ and $a=L_r$ , where $L_r$ is the  scale length of the radial modulation. To avoid spurious interferences, we require $L_r$ to be large enough to ensure that $1/(2\alpha a) \ll (\pi/\alpha)^{1/2}$. By substituting $\alpha$ with $\Psi''/2$, we obtain  the condition 
\begin{align}
L_r \gg (2\pi\Psi'')^{-1/2}\text{.}
\label{eq:ineq_L}
\end{align}
In the case of a constant-intensity line, and for $\delta\ll f_0$, the radial phase is~\cite{Smartsev:19}
\begin{align}
\Psi''= 3\frac{\delta k}{f_0^2} \frac{r^2}{R^2} + \mathcal{O}(r^4/R^4)\text{.}
\end{align}
Substituting this expression into Eq.~(\ref{eq:ineq_L}), we derive the condition:
\begin{align}
L_r\gg L_{r_0}=\frac{R}{r}\frac{f_0}{(6\pi k \delta)^{1/2}}\text{.}
\label{eq:Lr}
\end{align}
It is more informative to translate this radial scale length into an achievable longitudinal resolution along the focal line. For a constant-intensity line, this translation is straightforward, as we have the relation $z = f_0 + \delta r^2/R^2$. This leads to a longitudinal resolution
\begin{align}
L_z &\gg \frac{2\delta r L_{r_0}}{R^2} = \frac{f}{R} \left(\frac{2\delta}{3\pi k}\right)^{1/2}\text{.}
\label{eq:res_scale_length}
\end{align}
Interestingly, the achievable resolution is independent of the position along the focal line and depends only on the numerical aperture and the focal depth.
To validate this estimate, we performed a set of \texttt{Axiprop} simulations. We considered a radial intensity profile featuring two Gaussian holes located at $r_1=0.45R$ and $r_2=0.9R$,  described by  $E_i(r) =E_{i0}(r) - E_0\exp\left(-(r-r_i)^2/L_{ri}^2\right)$ where $i\in \{1,2\}$, $E_{i0}(r)$ is the radial profile without a hole, and $L_{ri} = L_r(r_i)$. Figure~\ref{fig:res} presents the simulation results for $R=50$~mm, $f_0\in\{200, 400\}$~mm, $\delta\in \{30,60\}$~mm and various values of $L_r\in\{1, 3,6,9\}L_{r_0}$.  

We observe that the hole depth is reduced and that fringes begin to appear for $L_r \leq 6 L_{r_0}$. Additionally, we find that $L_z$ increases linearly with $f_0$ and scales as $\delta^{-1/2}$ while remaining independent of the longitudinal position, as expected. Thus, Fig.~\ref{fig:res} validates the scaling law given by Eq.(\ref{eq:res_scale_length}). It also provides order-of-magnitude estimate for the achievable longitudinal resolution,
\begin{align}
L_z\approx 2.8 \frac{f_0}{R}\left(\frac{\delta}{k}\right)^{1/2}\text{.}
\label{eq:long_res}
\end{align}

\begin{figure}
    \centering

\includegraphics[width=\linewidth]{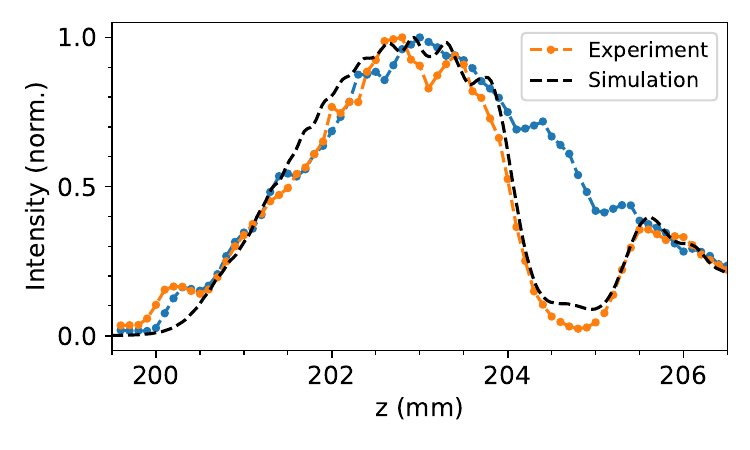}
    \caption{Experimental on-axis intensity profiles obtained with a smooth central, hole with (orange) and without (blue) a Gaussian ring hole defined by $L_1 = 1.82~\mathrm{mm}$ and $r_1 = 20.3~\mathrm{mm}$. Theoretical prediction from simulations is shown as dashed lines.}
    \label{fig:hole_exp}
\end{figure}
We now turn to an experimental validation of this theoretical resolution, using the same setup as in Sec~\ref{sec:ampl_exp}. Figure~\ref{fig:hole_exp} shows the recorded intensity profile for a radial profile featuring a smooth central hole, as described in Sec~\ref{sec:ampl_exp}, followed by a Gaussian ring with parameters $r_1 = 20.3~\mathrm{mm}$ and $L_1 = 1.82~\mathrm{mm}$. Each data point corresponds to the average of 10 images acquired with an exposure time of 200~ms, and a moving average with a window size of 3 data points was applied to smooth the profile.

As expected, a hole is observed in the plateau of the axiparabola.
By fitting the profile with a Gaussian function, we find that the hole has a width of 
$0.5~\mathrm{mm}$, which is of the same order of magnitude as the longitudinal resolution 
$L_z \approx 0.67~\mathrm{mm}$ given by Eq.~\ref{eq:long_res}. The theoretical curve shown in the figure indicates that the experimentally measured contrast is even slightly better than expected, likely related to optical aberrations. In principle, similar longitudinal modulations could be achieved using phase-only modulation.

\subsection{Oscillating profile}

Among the various forms of longitudinal intensity modulation, sinusoidal modulations have already attracted significant interest. They have been identified as a promising approach to boost X-ray emission in laser–plasma accelerators~\cite{Tomkus2020}, as well as to increase the electron energy through quasi-phase matching in laser-wakefield acceleration~\cite{PhysRevLett.112.134803,PhysRevLett.100.195001}. For the first application, the desired spatial period is of hundreds of microns while the second application is linked to the dephasing length going from hundreds of microns to few centimeters.
An efficient way to generate such longitudinal modulations is to imprint a radial intensity modulation on the beam prior to the optics that produce the quasi-Bessel beam~\cite{PhysRevLett.99.035001}.
Here, we build on this idea to determine, both numerically and experimentally, the achievable spatial frequencies.
\begin{figure}
    \centering
    \includegraphics[width=\linewidth]{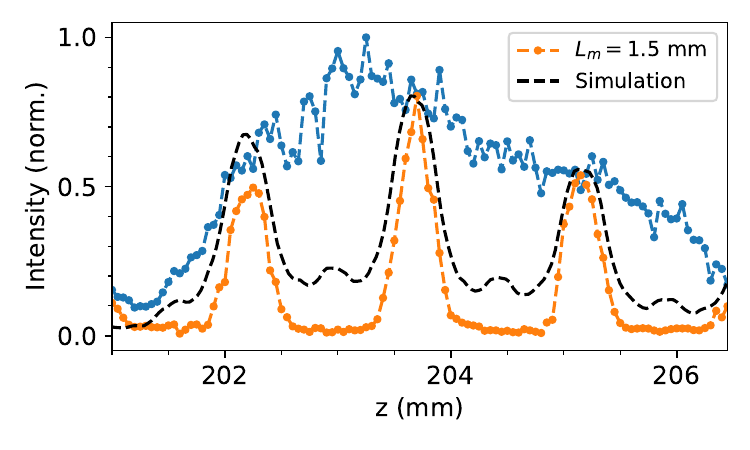}
    \caption{Experimental on-axis intensity profiles obtained with a smooth central hole, with (orange) and without (blue) sinusoidal modulation of spatial period $L_m = 1.5\,\text{mm}$. Theoretical prediction from simulations is shown as dashed lines.}
    \label{fig:oscil}
\end{figure}

As a preliminary step, we scanned several oscillation periods and found that well-contrasted oscillations persist for periods as short as $L_m = 1.5\,\text{mm}$, where $L_m$ denotes the period of the sinusoidal modulations we aim to achieve. This value is consistent with the longitudinal resolution $L_z$ determined in Sec.~\ref{sec:lmt_res}. The experimental on-axis intensity profile and the corresponding simulated curves for this minimum period are shown in Fig.~\ref{fig:oscil}.
While clearly resolved, the modulation, originally designed to be sinusoidal, exhibits sharper peaks and flatter troughs.
Interestingly, the contrast of the experimentally observed oscillations is higher than predicted by the simulations, in agreement with the trend already observed in Fig.~\ref{fig:hole_exp}.

These results demonstrate that millimeter-scale oscillations can be achieved with high contrast. Regarding applications, such millimeter-scale modulation appears too coarse for X-ray–emission enhancement schemes, which typically require sub-millimeter periods. However, it may be suited to quasi–phase matching in laser–wakefield accelerators.

\subsection{Segmented optics}

In the previous subsections, we saw that interference sets a fundamental limit on intensity-shaping resolution. 
Here, we show that segmented optics can overcome this limit, allowing the creation of intensity profiles otherwise inaccessible.
By dividing the optic into two or more concentric segments, with a temporal delay between adjacent segments at least equal to the pulse duration, temporal overlap between their contributions is avoided. This suppresses interference at the transitions, enabling steep variations without spurious oscillations.

Even with segmentation, the ultimate feature size remains constrained by diffraction and by the numerical aperture of each optic segment. These factors impose a fundamental resolution limit on the achievable longitudinal intensity structure. Considering, for example, a ring of light with inner radius $R_1$ and radial thickness $\sigma$, uniformly illuminated and focused by a parabolic optic of focal length $f_1$ the resulting on-axis intensity, obtained using Eq.~\eqref{eq:fresnel}, is
\begin{align}
I(z)\propto \mathrm{sinc}^2\!\left[\frac{k_0}{4}\bigl(\sigma^2+2R_1\,\sigma\bigr)\left(\frac{f_1-z}{f_1^2}\right)\right],
\label{eq:intensity_long}
\end{align}
leading to a standard deviation in the main lobe
\begin{align}
\sigma_z\approx\frac{2.1\lambda_0 f_1^2}{\pi(\sigma^2+2R_1\,\sigma)}=\frac{2.1\lambda_0 f_1^2}{S}.
\label{eq:res_long}
\end{align}
This shows that the resolution depends only on the illuminated area $S$ and the focal length $f_1$, and is independent of the ring’s radial position $R_1$. For a sufficiently large illuminated area, the resolution can thus be reduced to well below the millimeter scale.

We illustrate this approach by generating a flat-top intensity profile featuring a sharp spike, whose width cannot be achieved without segmented optics. 
This configuration uses an optic with two segments: a central parabolic region with $f_1 = 205\,\mathrm{mm}$ and $R_1 = 12\,\mathrm{mm}$, and an outer axiparabola defined by $f_0 = 200\,\mathrm{mm}$, $\delta = 30\,\mathrm{mm}$, and $R = 50\,\mathrm{mm}$. Additionally, amplitude modulation was applied to mitigate residual oscillations, with parameters $R_{\mathrm{hole}} = 18\,\mathrm{mm}$ and $L_{\mathrm{hole}} = 4\,\mathrm{mm}$.
\begin{figure}
    \centering
    \includegraphics[width=\linewidth]{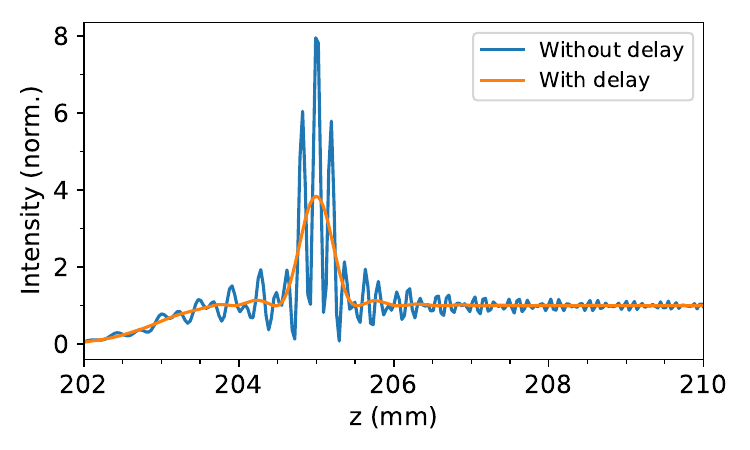}
    \caption{On-axis intensity profile obtained from simulations for a segmented axiparabola with and without radial delay.}
    \label{fig:segmented}
\end{figure}

The resulting on-axis intensity profiles are shown in Fig.~\ref{fig:segmented}. Without delay, the segmented design exhibits strong interference between contributions from different regions of the optic. Introducing a temporal delay between segments eliminates this overlap, achievable in practice through a longitudinal gap between the segments, and enables the formation of a sharp leading spike with a width of 0.16 mm -significantly smaller than both the experimental width and the theoretical resolution limit in Sec.~\ref{sec:lmt_res}.

This demonstrates that segmentation, combined with appropriate temporal delay, enables the generation of complex, high-resolution intensity features that cannot be achieved with a single continuous optic relying solely on amplitude or phase shaping. 
Such control over the longitudinal intensity profile opens the possibility of applying segmented optics to advanced laser–plasma concepts, including Trojan-horse injection~\cite{PhysRevLett.108.035001} and tailored plasma-density shaping for betatron-radiation generation~\cite{taphuoc:hal-00573304}.

\section{Conclusion and perspective}

In summary, we have identified the physical origin of the unwanted oscillations that commonly distort the longitudinal intensity profiles of quasi-Bessel beams. By analyzing beam formation in the Fresnel diffraction regime, we showed that these oscillations arise from the sharp truncation of the radial contributions to the Fresnel integral, which leads to interference between boundary terms and produces the characteristic modulations observed at the two ends of the focal line. Building on this understanding, we established two general strategies—amplitude shaping and phase-only tailoring—that reliably suppress these oscillations. 
While amplitude shaping provides a straightforward solution, only the phase-only approach is fully compatible with high-intensity beams.
Both approaches were validated through numerical simulations and dedicated experiments, showing excellent agreement with theoretical predictions.

Beyond removing distortions, we demonstrated that the same framework enables the controlled introduction of longitudinal modulations, 
and we analytically derived the fundamental resolution limit associated with these structures. 
We further showed that segmented optics with appropriate temporal delays can bypass this limit, allowing the realization of sharp, high-contrast features inaccessible to continuous optics.

These results provide a comprehensive methodology for designing extended focal lines with tailored longitudinal structure, ranging from fully smoothed profiles to engineered periodic or localized features. Such control opens promising avenues for high-intensity laser applications, including optimized laser-wakefield acceleration, enhanced betatron photon sources, and advanced plasma-shaping concepts. More broadly, the methods introduced here are applicable to any geometry relying on non-diffractive beams or axially extended foci, potentially enabling new regimes of interaction in ultrafast and high-field physics.

\bibliographystyle{apsrev4-2}
\bibliography{sample}

\end{document}